\documentclass[prb,showpacs,twocolumn,aps,superscriptaddress,floatfix]{revtex4}
\usepackage{amsmath}
\usepackage{amssymb}
\usepackage{bm}
\usepackage{graphicx}
\usepackage{color}
\usepackage{ulem}

\begin{document}

\title{AA-stacked bilayer graphene in an applied electric field: Tunable antiferromagnetism and coexisting exciton order parameter }

\author{R.S. Akzyanov}
\affiliation{Moscow Institute of Physics and Technology, Dolgoprudny, Moscow Region, 141700 Russia}
\affiliation{Institute for Theoretical and Applied Electrodynamics, Russian
Academy of Sciences, Moscow, 125412 Russia}
\affiliation{All-Russia Research Institute of Automatics, Moscow, 127055 Russia }

\author{A.O. Sboychakov}
\affiliation{Institute for Theoretical and Applied Electrodynamics, Russian
Academy of Sciences, Moscow, 125412 Russia}

\affiliation{CEMS, RIKEN, Saitama, 351-0198, Japan}

\author{A.V. Rozhkov}
\affiliation{Moscow Institute of Physics and Technology, Dolgoprudny, Moscow Region, 141700 Russia}
\affiliation{Institute for Theoretical and Applied Electrodynamics, Russian
Academy of Sciences, Moscow, 125412 Russia}
\affiliation{CEMS, RIKEN, Saitama, 351-0198, Japan}

\author{A.L. Rakhmanov}
\affiliation{Moscow Institute of Physics and Technology, Dolgoprudny, Moscow Region, 141700 Russia}
\affiliation{Institute for Theoretical and Applied Electrodynamics, Russian
Academy of Sciences, Moscow, 125412 Russia}
\affiliation{All-Russia Research Institute of Automatics, Moscow, 127055 Russia }

\affiliation{CEMS, RIKEN, Saitama, 351-0198, Japan}

\author{Franco Nori}
\affiliation{CEMS, RIKEN, Saitama, 351-0198, Japan}
\affiliation{Department of Physics, University of Michigan, Ann Arbor, MI 48109-1040, USA}

\begin{abstract}
We study the electronic properties of AA-stacked bilayer graphene in a
transverse electric field. The strong on-site Coulomb
repulsion stabilizes the antiferromagnetic order in such a system. The
antiferromagnetic order is suppressed by the transverse bias voltage, at
least partially. The inter-plane Coulomb repulsion and non-zero voltage
stabilize an exciton order parameter. The exciton order parameter coexists
with the antiferromagnetism and can be as large
as several tens of meV for realistic values of the bias voltage and
interaction constants. The application of a transverse bias voltage can be
used to control the transport properties of the bilayer.
\end{abstract}

\pacs{73.22.Pr, 73.22.Gk, 73.21.Ac}

%73.21.-b 	Electron states and collective excitations in multilayers,
%quantum wells, mesoscopic, and nanoscale systems (for electron states in
%nanoscale materials, see 73.22.-f)
%
%73.21.Ac 	Multilayers
%
%73.21.Cd 	Superlattices
%
%73.21.Fg 	Quantum wells
%
%73.21.Hb 	Quantum wires
%
%73.21.La 	Quantum dots
%
%73.22.-f 	Electronic structure of nanoscale materials and related
%systems
%
%73.22.Dj 	Single particle states
%
%73.22.Gk 	Broken symmetry phases
%
%73.22.Lp 	Collective excitations
%
%73.22.Pr 	Electronic structure of graphene

\maketitle

\section{Introduction}
The electronic properties of graphene are a subject of active
theoretical and experimental studies
\cite{castro_neto_review2009,chakraborty_review,meso_review}.
In addition to single-layer graphene, bilayer graphene also attracts
significant research attention. This interest is partly driven by the desire to
extend the family of graphene-like materials, and to create materials
with a controllable gap in the electronic spectrum.

The most studied form of bilayer is the AB (or Bernal) stacked bilayer graphene
(AB-BLG)~\cite{mccann2006,susp_bilayer2009,ABBLG,Mayorov_Sci2011,aleiner}.
The biased AB-BLG has a tunable
gap~\cite{biased_ab,Zhang_ab}. Excitons can
exist in the AB-BLG under certain
conditions~\cite{Fukidome2014,Nano_exciton}.

The AA-stacked bilayer graphene (AA-BLG) has received less
attention~\cite{aa_dft2008,spin-orbit2011,aa_adsorbtion2010,aa_optics_2010,aa first,
aa_experiment2008,borysiuk_aa2011,
our_preprint, Last_paper,our_preprint2}. However, samples of AA-BLG have recently been produced~\cite{aa_experiment2008,borysiuk_aa2011,aa first} and a detailed study of this system
becomes necessary. A significant feature of the AA-BLG is the perfect nesting of the hole and
electron Fermi surfaces. These degenerate Fermi surfaces are unstable with
respect of an arbitrarily weak electron interaction, and the AA-BLG becomes
an antiferromagnetic (AFM) insulator with a finite electron gap~\cite{our_preprint}. This electronic instability is strongest at zero doping, when the bands cross at the Fermi level.

An interesting phenomenon, which occurs in bilayer graphene
systems, is exciton
condensation~\cite{Eisenstein_nature,Coulomb_drag}.
In graphene bilayers, exciton condensation attracted
attention for both fundamental
reasons~\cite{Joglekar,Min,Perali,Rossi,Sokolik}
and possible applications in devices, including ultra-fast switches and
dispersionless field-effect transistors~\cite{BISFET}.

The purpose of this paper is to investigate the influence of a transverse
electric field on the properties of the AA-BLG. We show that such a field
can partially suppress the AFM order parameter. However, the degree of
suppression heavily depends on the effective value of the on-site Coulomb
repulsion. Moreover, the transverse bias stabilizes the exciton order parameter. Namely, we found that the exciton order parameter coexists with the AFM order if a transverse electric field is
applied. The exciton order is tuned by the voltage and tied to the AFM order.
Since the magnitude of the gap is sensitive to the transverse field, it
appears possible to control the transport properties of the bilayer with
the help of a transverse bias, which can be created by, e.g., a gate
electrode.

The paper is organized as follows. In section II we analyze
the single-electron part of our model. Within the tight-binding approach
we derive the degenerate electronic spectrum of the model.
In section III we consider the on-site and inter-site inter-plane
Coulomb repulsion using a mean-field theory. The electronic interaction
removes the degeneracy of the single electron spectrum creating a gap. We
found that the phase with coexisting AFM and exciton orders is the most
stable one. We obtain the equations for the order parameters and solve them
using both analytic and numerical methods.

\begin{figure}%[btp]
\centering
\includegraphics[width=0.95\columnwidth]{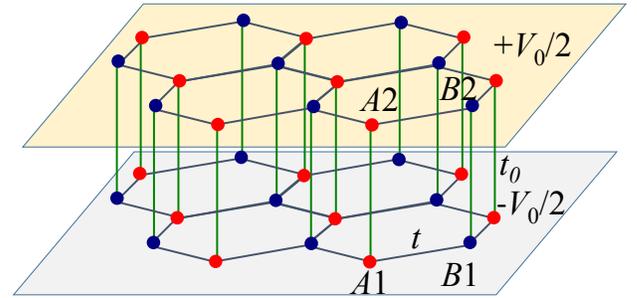}
\caption{(Color online) Crystal structure of the AA-stacked bilayer
graphene. The circles denote carbon atoms in the
${\cal A}$
(red) and
${\cal B}$
(blue) sublattices in the bottom (1), in grey, and top (2), in yellow, layers. The unit cell of
the AA-BLG consists of four atoms $A1$, $A2$, $B1$, and $B2$. The hopping
integrals $t$ and $t_0$ correspond to the in-plane and inter-plane
nearest-neighbor hopping. A transverse electrical voltage $V_0$ is applied to the planes.
\label{AABLG}}
\end{figure}

\section{Tight-binding Hamiltonian}\label{TBH}

The crystal structure of the AA-BLG is shown in Fig.~\ref{AABLG}. The
AA-BLG consists of two graphene layers, $1$ and $2$. Each carbon atom of the
upper layer is located above the corresponding atom of the lower layer.
Each layer consists of two triangular sublattices ${\cal A}$ and ${\cal
B}$. The elementary unit cell of the AA-BLG contains four carbon atoms
$A1$, $A2$, $B1$, and $B2$.

We write the single-particle tight-binding Hamiltonian of the AA-BLG in the form
\begin{eqnarray}\label{H0}
H_0&=&-t\sum_{\langle\mathbf{nm}\rangle i\sigma}\left(d^{\dag}_{\mathbf{n}i{\cal A}\sigma}
d^{\phantom{\dag}}_{\mathbf{m}i{\cal B}\sigma}+H.c.\right)+
\nonumber\\
&&t_0\sum_{\mathbf{n}a\sigma}\left(d^{\dag}_{\mathbf{n}1a\sigma}d^{\phantom{\dag}}_{\mathbf{n}2a\sigma}+H.c.\right)
+\\
&&\frac {V_0}{2}\sum_{\mathbf{n}a\sigma}\left(d^{\dag}_{\mathbf{n}1a\sigma}d^{\phantom{\dag}}_{\mathbf{n}1a\sigma}-d^{\dag}_{\mathbf{n}2a\sigma}d^{\phantom{\dag}}_{\mathbf{n}2a\sigma}\right) .\nonumber
\end{eqnarray}
Here $d^{\dag}_{\mathbf{n}ia\sigma}$
and
$d^{\phantom{\dag}}_{\mathbf{n}ia\sigma}$
are the creation and annihilation operators of an electron with spin
projection $\sigma$ in the layer
$i=1,2$
on the sublattice
$a={\cal A},{\cal B}$
at the position
$\mathbf{n}$, and
$\langle ...\rangle$
denotes a nearest-neighbor pair inside a layer. The amplitude $t$ ($t_0$) in
Eq.~\eqref{H0} describes the in-plane (inter-plane) nearest-neighbor
hopping, $V_0$ is the voltage applied perpendicular to the layers. We
assume that
$V_0 \ll t$,
which corresponds to typical experimental
conditions~\cite{biased_ab,Zhang_ab}.
For calculations, we use the values of the hopping integrals
$t=2.57$\,eV, $t_0=0.36$\,eV, computed by the DFT method for multilayer carbon systems in Ref.~\onlinecite{Charlier}.

After diagonalizing the Hamiltonian \eqref{H0} we obtain four bands
$\varepsilon^{(s)}_{0\mathbf{k}}$ ($s=1,\dots,4$), which can be written as
\begin{eqnarray}
\label{E0k}
&&\varepsilon^{(1)}_{0\mathbf{k}}=-\sqrt{t_0^2+\frac {V_0^2}4}-t\zeta_{\bf k}\,,\,\,
\varepsilon^{(2)}_{0\mathbf{k}}=-\sqrt{t_0^2+\frac {V_0^2}4}+t\zeta_{\bf k}\,,\nonumber\\
&&\varepsilon^{(3)}_{0\mathbf{k}}=\sqrt{t_0^2+\frac {V_0^2}4}-t\zeta_{\bf k}\,,\,\,\,\,\,
\varepsilon^{(4)}_{0\mathbf{k}}=\sqrt{t_0^2+\frac {V_0^2}4}+t\zeta_{\bf k}\,,
\end{eqnarray}
where
\begin{equation}
\zeta_{\bf k} = |f_{\bf k}|\,,\;\;f_{\mathbf{k}}\!=\!1+2\exp\!\left(\frac{3ik_xa_0}{2}\right)\cos\!\left(\!\!\frac{\sqrt{3}k_ya_0}{2}\!\!\right),
\end{equation}
and $a_0=1.42$\,{\AA} is the in-plane carbon-carbon distance.
The bands $s=2$ and $s=3$ cross the Fermi level near the Dirac points
$\mathbf{K}=2\pi (\sqrt{3},\,1 )/(3\sqrt{3}a_0)$
and $\mathbf{K}'=2\pi (\sqrt{3},\,-1 )/(3\sqrt{3}a_0)$. As it follows from Eqs.~\eqref{E0k}, the band $s=2$ is electron-like, while the band $s=3$ is hole-like.
The band $s=1$ lies below and the band $s=4$ lies above the Fermi energy and, consequently, they do not form a Fermi surface.

In contrast to the Bernal stacking (where the bias voltage opens a gap at the Fermi level~\cite{biased_ab,Zhang_ab}), the application of a transverse bias voltage does not qualitatively change the spectrum of the AA-BLG. The Fermi surface is given by the equation
$\zeta_{\bf k}=t^{-1}\sqrt{t_0^2+V_0^2/4} = \zeta_0$.
Since
$\zeta_0\ll1$,
we can expand the function
$\zeta_{\bf k}$
near the Dirac points and find that the Fermi surface consists of two
circles with radius
$k_{r}=2\zeta_0/(3a_0)$.

One of the most important features of this tight-binding band structure is
that the Fermi surfaces of both bands coincide. That is, the electron and
hole components of the Fermi surface are perfectly nested. This property
of the Fermi surface is quite robust against changes in the tight-binding
Hamiltonian. It survives even if longer-range hoppings are taken into
account, or a system with two non-equivalent layers is considered (e.g.,
the single-side hydrogenated graphene~\cite{sshg}).
However, the electron interactions can remove the degeneracy in the spectrum, creating a finite
gap~\cite{our_preprint}.

\section{Electron-electron interaction}\label{CAFM}

The electronic spectrum changes drastically when considering the Coulomb
interaction. To study the effects of this interaction on the electronic
properties of our system, we use the following Hubbard-like Hamiltonian
\begin{equation}\label{U}
H_{\textrm{int}}=H_{\textrm{U}}+H_{\textrm{V}}\,.
\end{equation}
The first term, $H_{\textrm{U}}$, is the on-site Coulomb repulsion between the electrons,
\begin{equation}\label{HU}
H_{\textrm{U}}=U_0\sum_{\mathbf{n}ia}
\left(n_{\mathbf{n}ia\uparrow}-\frac{1}{2}\right)\!\!\left(n_{\mathbf{n}ia\downarrow}-\frac{1}{2}\right)\,,
\end{equation}
where $n_{\mathbf{n}ia\sigma}=d^{\dag}_{\mathbf{n}ia\sigma}d^{\phantom{\dag}}_{\mathbf{n}ia\sigma}$ is the operator of the occupation number. The second term, $H_{\textrm{V}}$, describes the nearest-neighbor Coulomb repulsion. It has a form
\begin{eqnarray}\label{HV}
H_{\textrm{V}}&=&U_{12}\sum_{\scriptstyle\mathbf{n}a\atop\scriptstyle\sigma\sigma'}\left(n_{\mathbf{n}1a\sigma}-\frac{1}{2}\right)\!\!\left(n_{\mathbf{n}2a\sigma'}-\frac{1}{2}\right)+\nonumber\\
&&U_{ab}\!\sum_{\scriptstyle\langle\mathbf{nm}\rangle\atop\scriptstyle i\sigma\sigma'}\left(n_{\mathbf{n}i{\cal A}\sigma}-\frac{1}{2}\right)\!\!\left(n_{\mathbf{m}i{\cal B}\sigma'}-\frac{1}{2}\right)\!,
\end{eqnarray}
where the first term is the nearest-neighbor interaction between the electrons in different layers, while the second term describes the in-plane nearest-neighbor interaction. The terms $1/2$ in the brackets in Eqs.~\eqref{HU} and~\eqref{HV} are added to keep the chemical potential corresponding to the half-filling (zero doping) equal to zero.

The value of the electron-electron interaction in graphene is relatively
strong. According to DFT calculations~\cite{U69}, the on-site repulsion
energy, $U_0$, is about 9--10\,eV, while the in-plane inter-site
repulsion, $U_{ab}$, is about 5--6\,eV. The nearest-neighbor inter-plane
interaction in the bilayer graphene is unknown. We can estimate it as
$U_{12}\approx U_{ab}\,a_0/c\approx2.5$\,eV, where $c=3.35$\,{\AA} is the
distance between the layers. It is commonly accepted that the mean-field
calculations overestimate the resulting value of the antiferromagnetic (AFM)
order parameter driven by the electron-electron interaction. In addition,
the long range Coulomb interaction can effectively
reduce~\cite{Effective_onsite} the on-site repulsion energy $U_0$.
Keeping all this in mind, we use for further estimates the values of $U_0$, $U_{ab}$, and $U_{12}$ smaller than those obtained in the DFT calculations. We will use $U_0/t \simeq$2--3.5, $U_{ab}/t\simeq$1--2, and $U_{12}/t\simeq$0.5--1.

\subsection{Mean-field equations}\label{CAFMA}

We analyze the properties of the total Hamiltonian $H=H_{0}+H_{\textrm{int}}$ in
the mean-field approximation. It was shown previously for zero bias voltage
that the on-site Coulomb repulsion stabilizes the AFM ground state in the
AA-BLG~\cite{our_preprint,Last_paper,our_preprint2}. We will show below
that the AFM order also exists for
$V_0\neq0$.
We fix the spin quantization $z$-axis perpendicular to the layers in the $xy$ plane.
In this case the AFM order parameter can be written as
\begin{eqnarray}
\Delta_{\textrm{AFM}}^{ia}=\frac{U_0}{2}\left(\left\langle
d^{\dag}_{\mathbf{n}ia\uparrow}d^{\phantom{\dag}}_{\mathbf{n}ia\uparrow}\right\rangle
-\left\langle
d^{\dag}_{\mathbf{n}ia\downarrow}d^{\phantom{\dag}}_{\mathbf{n}ia\downarrow}
\right\rangle\right)\,,\\
%%%%%%%%%%%%%%%%%%%%%%%%%%%%%%%%%%%%%%%%%%%%%%%%%%
\label{GtypeDelta}
%%%%%%%%%%%%%%%%%%%%%%%%%%%%%%%%%%%%%%%%%%%%%%%%%%
\Delta_{\textrm{AFM}}^{1\cal{A}}=
\Delta_{\textrm{AFM}}^{2\cal{B}}=-\Delta_{\textrm{AFM}}^{1\cal{B}}=-\Delta_{\textrm{AFM}}^{2\cal{A}}\equiv\Delta_{\textrm{AFM}}\,,
\end{eqnarray}
and the $\Delta_{\textrm{AFM}}$ is real. Such an AFM order, when  the spin at any given site is
antiparallel to spins at all its nearest-neighbor sites, is referred as G-type AFM. In the mean-field approximation, the on-site interaction Hamiltonian, $H_{\text{U}}$, takes the form
\begin{eqnarray}
H^{\rm MF}_{\text{U}}\!\!&=&\!-\frac{{\cal N}U_0\Delta n^2}{4}+\frac{U_0\Delta n}{4}\sum_{\mathbf{n}a\sigma}\left(n_{\mathbf{n}1a\sigma}-n_{\mathbf{n}2a\sigma}\right)+\nonumber\\
&&\frac{4{\cal N}\Delta_{\textrm{AFM}}^2}{U_0}
-\!\sum_{\mathbf{n}ia}\!\Delta_{\textrm{AFM}}^{ia}\!\left(n_{\mathbf{n}ia\uparrow}-n_{\mathbf{n}ia\downarrow}\!\right)\!,\label{UMF}
\end{eqnarray}
where $\Delta n=\sum_{\sigma}(\langle n_{\mathbf{n}1a\sigma}\rangle-\langle n_{\mathbf{n}2a\sigma}\rangle)$ is the difference in the electron densities in two graphene layers induced by the applied voltage $V_0$ and ${\cal N}$ is the number of unit cells in the sample.

Let us consider now the inter-site part of the interaction. The Hamiltonian
$H_{\text{V}}$ can produce several order parameters in the system.
However, for zero-bias voltage all of them compete with the
antiferromagnetism and only the antiferromagnetic order parameter survives, because
$U_0$ is the strongest interaction constant.
A nonzero bias voltage breaks the symmetry between two graphene layers. In
this case, there exists an order parameter driven by the inter-layer interaction,
which coexists with antiferromagnetism. An analysis based on symmetry
considerations, similar to that presented in
Ref.~\onlinecite{our_preprint}, shows that this order parameter should have a form
\begin{eqnarray}
\Delta_{\textrm{exc}}^{a}&=&\frac{U_{12}}{2}\left(\left\langle d^{\dag}_{\mathbf{n}1a\uparrow}d^{\phantom{\dag}}_{\mathbf{n}2a\uparrow}\right\rangle-
\left\langle d^{\dag}_{\mathbf{n}1a\downarrow}d^{\phantom{\dag}}_{\mathbf{n}2a\downarrow}\right\rangle\right)\,,\\
%%%%%%%%%%%%%%%%%%%%%%%%%%%%%%%%%%%%%%%%%%%%%%%%%%
\label{ExcitonSymmetry}
%%%%%%%%%%%%%%%%%%%%%%%%%%%%%%%%%%%%%%%%%%%%%%%%%%
\Delta_{\textrm{exc}}^{\cal{A}}&=&-\Delta_{\textrm{exc}}^{\cal{B}}\,\equiv\,\Delta_{\textrm{exc}}\,,
\end{eqnarray}
and the $\Delta_{\textrm{exc}}$ is real. This order parameter corresponds to
the bound state of the electron and the hole in different layers. We call
it the exciton order parameter.
The mean-field expression for the
inter-site part of the Hamiltonian has the following form
\begin{eqnarray}
%%%%%%%%%%%%%%%%%%%%%%%%%%%%%%%%%%%%%%%%%%%%%%%%%%
\label{UMF}
%%%%%%%%%%%%%%%%%%%%%%%%%%%%%%%%%%%%%%%%%%%%%%%%%%
H^{\rm MF}_{\textrm{V}}\!&=&\!{\cal N}\left[-\frac{(3U_{ab}-U_{12})\Delta n^2}{2}+\frac{4\Delta_{\textrm{exc}}^2}{U_{12}}\right]+\\
&&\frac{(3U_{ab}-U_{12})\Delta n}{2}\sum_{\mathbf{n}a\sigma}\left(n_{\mathbf{n}1a\sigma}-n_{\mathbf{n}2a\sigma}\right)-\nonumber\\
&&\!\sum_{\mathbf{n}a\sigma}\!\left[\Delta_{\textrm{exc}}^{a}\!\left(d^{\dag}_{\mathbf{n}1a\uparrow}d^{\phantom{\dag}}_{\mathbf{n}2a\uparrow}-
d^{\dag}_{\mathbf{n}1a\downarrow}d^{\phantom{\dag}}_{\mathbf{n}2a\downarrow}\right)+\!H.c.\!\right]\!.\nonumber
\end{eqnarray}

We introduce the four-component spinor
\begin{eqnarray}
\psi^{\dag}_{\mathbf{k}\sigma}
=
(
	d^{\dag}_{\mathbf{k}1\cal{A}\sigma},
	d^{\dag}_{\mathbf{k}2\cal{A}\sigma},
	d^{\dag}_{\mathbf{k}1\cal{B}\sigma},
	d^{\dag}_{\mathbf{k}2\cal{B}\sigma}
)\,.
\end{eqnarray}
In terms of this spinor, the mean field Hamiltonian
\begin{equation}\label{MF}
H^{\rm MF}=H_{0}+H^{\rm MF}_{\rm{U}}+H^{\rm MF}_{\rm{V}}
\end{equation}
can be written as
\begin{eqnarray}
\label{HtotM}
H^{\rm MF}
=
{\cal N}E_0 + \sum_{\mathbf{k}\sigma}
	\psi^{\dag}_{\mathbf{k}\sigma}(\hat{H}_{0\mathbf{k}}+\hat{\Delta}_{\sigma})
			\psi^{\phantom{\dag}}_{\mathbf{k}\sigma}\,,
\end{eqnarray}
where $E_0$ is a $c$-number
\begin{equation}
E_0=-\frac{(U_0+6U_{ab}-2U_{12})\Delta n^2}{4}+\frac{4\Delta_{\textrm{AFM}}^2}{U_0}+\frac{4\Delta_{\textrm{exc}}^2}{U_{12}}\,,
\end{equation}
and $\hat{H}_{0\mathbf{k}}$ and $\hat{\Delta}_{\sigma}$ are the $4\times4$ matrices
\begin{equation}\label{Hk}
\hat{H}_{0\mathbf{k}}=-\left(
\begin{matrix}
-V/2&t_0&tf_{\bf k}&0\cr
t_0&V/2&0&tf_{\bf k}\cr
tf_{\bf k}^{*}&0&-V/2&t_0\cr
0&tf_{\bf k}^{*}&t_0&V/2\cr
\end{matrix}\right)\!,
\end{equation}
\begin{equation}\label{DeltaMatr}
\hat{\Delta}_{\uparrow}=\left(
\begin{matrix}
\Delta_{\textrm{AFM}}&\Delta_{\textrm{exc}}&0&0\cr
\Delta_{\textrm{exc}}&-\Delta_{\textrm{AFM}}&0&0\cr
0&0&-\Delta_{\textrm{AFM}}&-\Delta_{\textrm{exc}}\cr
0&0&-\Delta_{\textrm{exc}}&\Delta_{\textrm{AFM}}\cr
\end{matrix}\right)\!,\;\hat{\Delta}_{\downarrow}=-\hat{\Delta}_{\uparrow}.
\end{equation}
In Eq.~\eqref{Hk} the quantity $V$ is the effective bias voltage given by the relation
\begin{equation}\label{Veff}
V=V_0+\alpha\Delta n\,,\,\,\,\;\;\alpha=\frac{U_0+6U_{ab}-2U_{12}}{2}>0\,.
\end{equation}
This equation describes the screening of the applied voltage due to the
electron-electron interaction. Indeed, since $\Delta n<0$ for $V_0>0$, we
have $V<V_0$.  For the parameters $U_0$, $U_{ab}$, and $U_{12}$
under study the constant $\alpha$ can be estimated as 3.5$t$--7$t$.

The mean-field spectrum is obtained by the diagonalization of two $4\times4$ matrices in Eq.~\eqref{HtotM}. It consists of four bands doubly-degenerate with respect to spin
\begin{eqnarray}\label{Ek}
\varepsilon^{(1)}_{\mathbf{k}}&=&-\sqrt{A_{\mathbf{k}}+2B_{\mathbf{k}}}\,,\;
\varepsilon^{(2)}_{\mathbf{k}}=-\sqrt{A_{\mathbf{k}}-2B_{\mathbf{k}}}\,,\nonumber\\
\varepsilon^{(3)}_{\mathbf{k}}&=&\sqrt{A_{\mathbf{k}}-2B_{\mathbf{k}}}\,,\;\;\;\;
\varepsilon^{(4)}_{\mathbf{k}}=\sqrt{A_{\mathbf{k}}+2B_{\mathbf{k}}}\,,
\end{eqnarray}
where
\begin{eqnarray}\label{AkBk}
A_{\mathbf{k}}&=&\Delta_{\textrm{AFM}}^2+\Delta_{\textrm{exc}}^2+t^2\zeta_{\mathbf{k}}^2+t_0^2+\frac {V^2}4\,,\\
B_{\mathbf{k}}&=&\sqrt{\left[-\Delta_{\textrm{exc}}t_0+\Delta_{\textrm{AFM}}\frac V2\right]^2+t^2\zeta_{\mathbf{k}}^2\left[t_0^2+\frac {V^2}4\right]}\,.\nonumber
\end{eqnarray}
The full gap in the spectrum $\Delta$ is defined as
$\Delta=\min_{\mathbf{k}}\left(\varepsilon^{(3)}_{\mathbf{k}}-\varepsilon^{(2)}_{\mathbf{k}}\right)$/2.
It relates to the AFM and exciton order parameters as
\begin{eqnarray}
%%%%%%%%%%%%%%%%%%%%%%%%%%%%%%%%%%%%%%%%%%%%%%%%%%
\label{Full_gap}
%%%%%%%%%%%%%%%%%%%%%%%%%%%%%%%%%%%%%%%%%%%%%%%%%%
\Delta
=
\frac {
	2 \Delta_{\textrm{AFM}}t_0+\Delta_{\textrm{exc}} V
      }
      {
	\sqrt{4 t_0^2+V^2}
      }\,.
\end{eqnarray}

To determine the values of the order parameters $\Delta_{\textrm{AFM}}$ and  $\Delta_{\textrm{exc}}$  we should minimize the grand potential
$\Omega$. The grand potential per unit cell is
\begin{equation}\label{Omega}
\Omega=E_0-2T\!\sum_{s=1}^{4}\!\int\!\frac{d\mathbf{k}}{V_{\text{BZ}}}\ln\left[1+e^{-\varepsilon^{(s)}_{\mathbf{k}}/T}\right]\,,
\end{equation}
where $V_{\text{BZ}}$ is the volume of the first Brillouin zone.

To calculate the integrals over the Brillouin zone, it is convenient to introduce
the density of states
\begin{equation}\label{ro}
\rho_0(\zeta)=\!\int\!\frac{d\mathbf{k}}{V_{\text{BZ}}}\delta(\zeta-\zeta_{\mathbf{k}})\,.
\end{equation}
This function is non-zero only if $0<\zeta<3$.
It is related to the single layer graphene density of states
$\rho_{\textrm{gr}}(\zeta)$
as
$\rho_{\textrm{gr}}(\zeta)=\rho_{0}(\zeta)/t$
(see Ref.~\onlinecite{castro_neto_review2009}).

Minimization of $\Omega$ with respect to $\Delta_{\textrm{AFM}}$ and  $\Delta_{\textrm{exc}}$ gives the equations
\begin{eqnarray}\label{First_equation}
\frac{4\Delta_{\textrm{AFM}}}{U_0}&=&\int\limits_{0}^{3}\!\!d\zeta\,\rho_0(\zeta)\!\!
\left[\Delta_{\textrm{AFM}}+\frac {V}{2} \theta(\zeta)\right]\!F\left(\!\varepsilon^{(1)}(\zeta)\!\right)+\nonumber\\
&&\!\int\limits_{0}^{3}\!\!d\zeta\,\rho_0(\zeta)\!\!\left[\!\Delta_{\textrm{AFM}}-\frac {V}{2}\theta(\zeta)\!\right]\!F\left(\!\varepsilon^{(2)}(\zeta)\!\right),
\end{eqnarray}
\begin{eqnarray}
%%%%%%%%%%%%%%%%%%%%%%%%%%%%%%%%%%%%%%%%%%%%%%%%%%
\label{Second_equation}
%%%%%%%%%%%%%%%%%%%%%%%%%%%%%%%%%%%%%%%%%%%%%%%%%%
\frac{4\Delta_{\textrm{exc}}}{U_{12}}
&=&
\int\limits_{0}^{3}\!\!d\zeta\,\rho_0(\zeta)\!
\left[\Delta_{\textrm{exc}}-t_0\theta(\zeta)\right]F\left(\varepsilon^{(1)}(\zeta)\right)+\nonumber\\
&&\int\limits_{0}^{3}\!\!d\zeta\,\rho_0(\zeta)\!\left[\Delta_{\textrm{exc}}+t_0\theta(\zeta)\right]F\left(\varepsilon^{(2)}(\zeta)\right),
\end{eqnarray}
where
\begin{eqnarray}
F(\varepsilon)
=
\frac{f(-\varepsilon)-f(\varepsilon)}{\varepsilon},\;\;\;\,
f(\varepsilon)=\frac{1}{e^{{\scriptstyle \varepsilon}/{\scriptstyle T}}+1}\,,
\nonumber \\
\theta(\zeta)
=
\frac{
	2 \Delta_{\textrm{exc}}t_0-\Delta_{\textrm{AFM}} V
     }
     {
	\sqrt{
		(2 \Delta_{\textrm{exc}}t_0-\Delta_{\textrm{AFM}} V)^2
		+
		t^2(4 t_0^2+{V^2})\zeta^2
	     }
     }\,,
\end{eqnarray}
and
$\varepsilon^{(s)}(\zeta)$
are given by Eqs.~\eqref{Ek}
and~\eqref{AkBk},
in which
$\zeta_{\mathbf{k}}$
is replaced by $\zeta$.

Equations~\eqref{First_equation} and~\eqref{Second_equation} define the AFM and
exciton order parameters as functions of the effective bias voltage $V$. In
order to find the dependencies of $\Delta_{\textrm{AFM}}$ and
$\Delta_{\textrm{exc}}$ on the applied voltage $V_0$ we should use
Eq.~\eqref{Veff}. To find the charge imbalance between two graphene layers
$\Delta n$,
we apply the Hellman-Feynman
theorem~\cite{Feynman}
\begin{eqnarray}
\Delta n=2\left\langle\frac{\partial H^{\rm MF}}{\partial V}\right\rangle=2\frac {\partial E}{\partial V}\,,
\end{eqnarray}
where $E$ is the energy of the system per unit cell. It can be written as
\begin{eqnarray}\label{Etot}
E=E_0+\!\sum_{s=1}^{4}\!\int\limits_{0}^{3}\!\!d\zeta\,\rho_0(\zeta)\,\varepsilon^{(s)}(\zeta)f\left(\varepsilon^{(s)}(\zeta)\right).
\end{eqnarray}
As a result, the expression for the renormalized bias $V$ takes the following form
\begin{eqnarray}\label{effective_bias}
V=V_0+2\alpha\frac{\partial E(V)}{\partial V}\,.
\end{eqnarray}
This equation, together with Eqs.~\eqref{First_equation}
and~\eqref{Second_equation},
define the AFM and exciton order parameters as functions of the applied
voltage.

\subsection{Analytical results}

In this subsection we obtain the solution of
Eqs.~\eqref{First_equation},
\eqref{Second_equation},
and~\eqref{effective_bias}
in the limits
$\Delta_{\textrm{exc}}\ll\Delta_{\textrm{AFM}}\ll t_0$
and
$T=0$.
When these conditions hold, the functions
$\varepsilon^{(1,2)}(\zeta)$
and
$\theta(\zeta)$
become
\begin{eqnarray}
\varepsilon^{(1)}(\zeta)&\cong&-\sqrt{\Delta_{\textrm{AFM}}^2+t^2(\zeta-\zeta_0)^2}\,,\nonumber\\
\varepsilon^{(2)}(\zeta)&\cong&-t(\zeta+\zeta_0)\,,\\
\theta(\zeta)
&\cong&
\frac{
	2 \Delta_{\textrm{exc}}t_0-\Delta_{\textrm{AFM}} V
     }
     {
	2 t^2\zeta_0\zeta
     }\,,\nonumber
\end{eqnarray}
where
\begin{eqnarray}
\zeta_0=\frac{\sqrt{t_0^2+V^2/4}}t.
\end{eqnarray}
Substituting
$\varepsilon^{(1,2)}(\zeta)$
with $\Delta_{\textrm{AFM}}=0$ in Eq.~\eqref{Etot}, we obtain the following relation between $V$ and $V_0$
\begin{equation}
%%%%%%%%%%%%%%%%%%%%%%%%%%%%%%%%%%%%%%%%%%%%%%%%%%
\label{VvsV0}
%%%%%%%%%%%%%%%%%%%%%%%%%%%%%%%%%%%%%%%%%%%%%%%%%%
V=\frac{V_0t}{t+{C\alpha \zeta_0 }}\,,\;\;
C=\left.\frac{\partial\rho_0(\zeta)}{\partial\zeta}\right|_{\zeta\rightarrow0}\!\!\!\simeq0.37\,.
\end{equation}
For realistic parameter values, the renormalized bias voltage $V$
depends almost linearly on $V_0$. Taking
$\alpha=10$\,eV,
we obtain from
Eq.~\eqref{VvsV0} that
$V=0.83V_0$,
if
$V_0\ll t_0$.
The numerical analysis shows that the estimation
$V\approx V_0$
becomes even better for larger values of
$\Delta_{\textrm{AFM}}/t$
and
$\Delta_{\textrm{exc}}/t$.

The analytical expressions for the order parameters are derived in the Appendix. The results can be rewritten as
\begin{eqnarray}\label{ExcitonGap}
\Delta_{\textrm{AFM}}\!=\!2t\!\sqrt{\!\zeta_0(\!3\!-\!\zeta_0\!)}\!\exp\!\left(\!-\frac{\frac{4t}{U_0}-\eta_1(\zeta_0)-\frac{\eta_2(\zeta_0)V^2}{4t^2\zeta_0^2}}{2\rho_0(\zeta_0)\frac{t_0^2}{t^2\zeta_0^2}}\!\right)\!,\nonumber\\
\Delta_{\textrm{exc}}\!=\!\Delta_{\textrm{AFM}}\!\frac{V}{2t_0}\frac{\frac{4t}{U_0}-\eta_1(\zeta_0)-\eta_2(\zeta_0)\zeta_0}{\frac{4t}{U_{12}}\!-\!\frac{4t}{U_0}\frac{t^2\zeta_0^2}{t_0^2}\!+\!\frac{V^2(\eta_1(\zeta_0)\!+\!\eta_2(\zeta_0)\zeta_0)}{4t_0^2}},\quad
\end{eqnarray}
where $\eta_1(\zeta_0)$ and $\eta_2(\zeta_0)$ are defined in the Appendix by
Eqs.~\eqref{integrals}. We see that $\Delta_{\textrm{exc}}$ is
proportional to the $\Delta_{\textrm{AFM}}$. When $V_0\ll t_0$, the exciton
order parameter depends linearly on $V_0$.

\subsection{Gap suppression by the transverse bias}

\begin{figure}
\centering
\includegraphics[width=0.95\columnwidth]{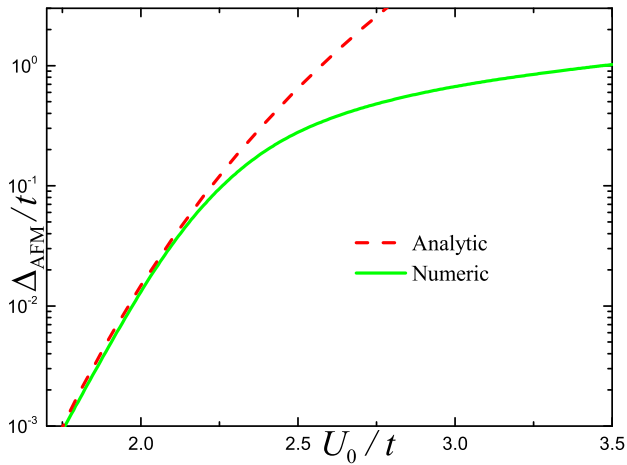}
\caption{(Color online) AFM order parameter $\Delta_{\textrm{AFM}}$ versus the
on-site interaction $U_0$, for zero bias $V_0=0$.}
\label{Delta012}
\end{figure}

The total gap in the spectrum is given by Eq.~\eqref{Full_gap}. It
coincides with the AFM order parameter if the bias voltage is zero. The
dependence of the gap on the ratio $U_0/t$ for zero bias is shown in Fig.~\ref{Delta012}. The
analytical expression Eq.~\eqref{ExcitonGap}
works well for $U_0\leq 2.3t$.

\begin{figure}
\centering
\includegraphics[width=0.95\columnwidth]{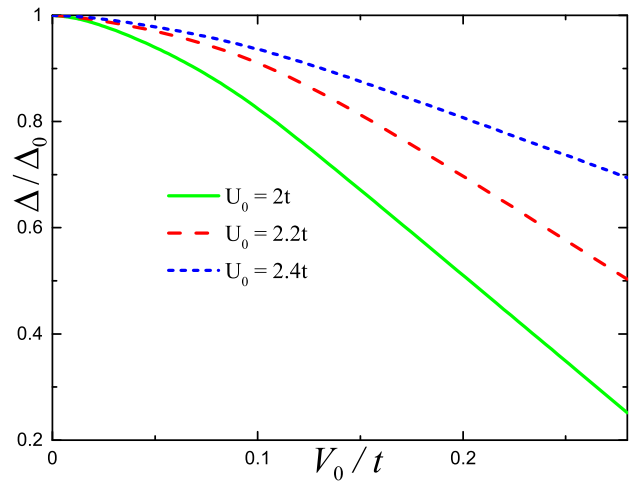}
\caption{(Color online) Full gap $\Delta$ versus the
applied bias $V_0$, for different values of the on-site Coulomb repulsion $U_0$. The value $\Delta_0$ is equal to the full gap if $V_0=0$, that is, it is
the AFM gap $\Delta_{\textrm{AFM}}$ for zero bias.}
\label{Fias}
\end{figure}

If
$V_0\neq0$,
the exciton order parameter becomes non-zero. The full gap $\Delta$,
however, decreases when $V_0$ increases. The dependence of the full gap
$\Delta$ on $V_0$ calculated for three different values of $U_0$ is shown
in Fig.~\ref{Fias}. As it follows from this figure, the gap suppression is
stronger for smaller $U_0$.

We consider here only the case of zero temperature. In this case
the full gap never reaches zero for realistic values of the applied
voltage. At finite temperatures, however, it can be fully suppressed by the
bias voltage. This makes it possible to observe a voltage-driven
metal-insulator transition.


\begin{figure}
\centering
\includegraphics[width=0.95\columnwidth]{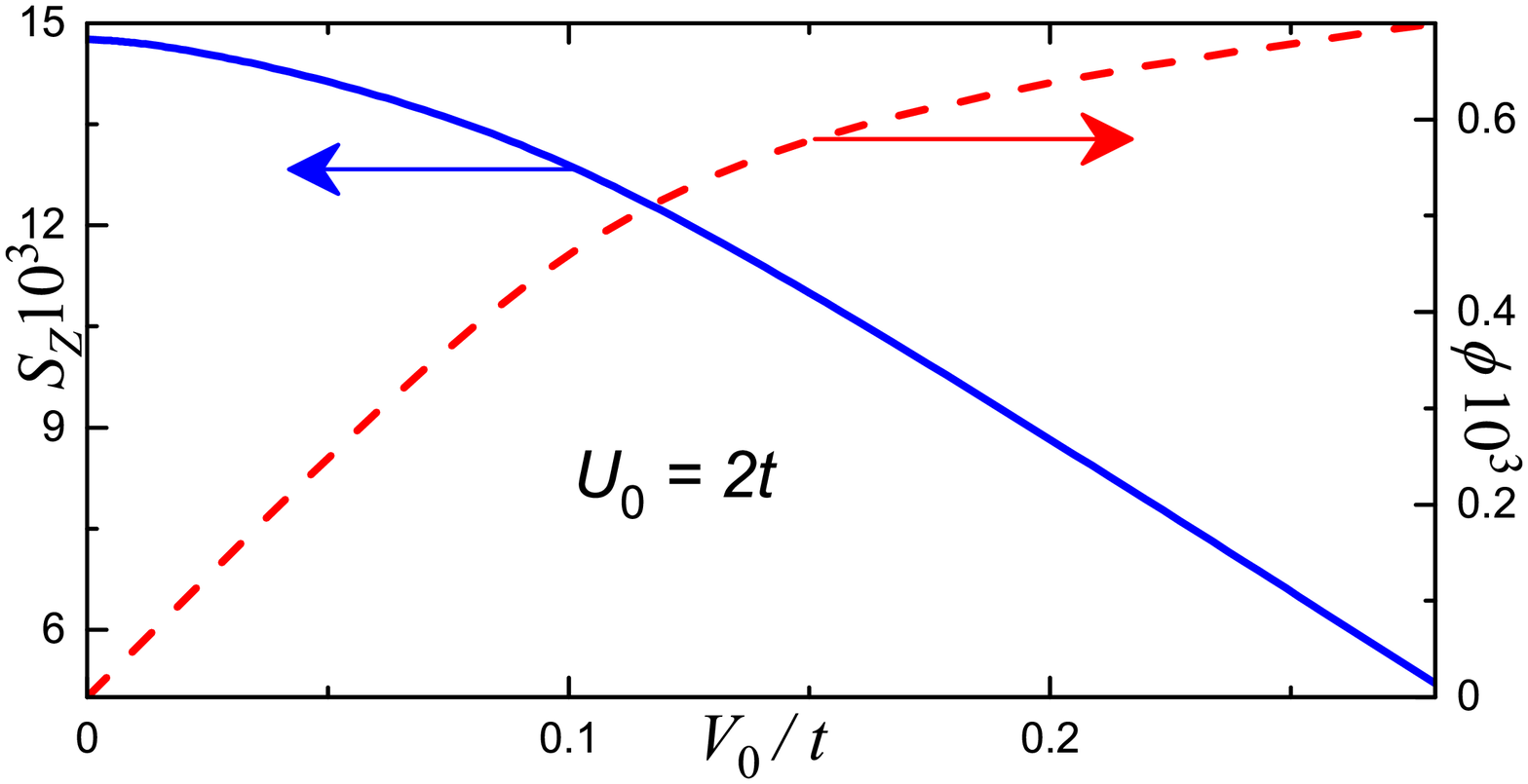}\\
\includegraphics[width=0.95\columnwidth]{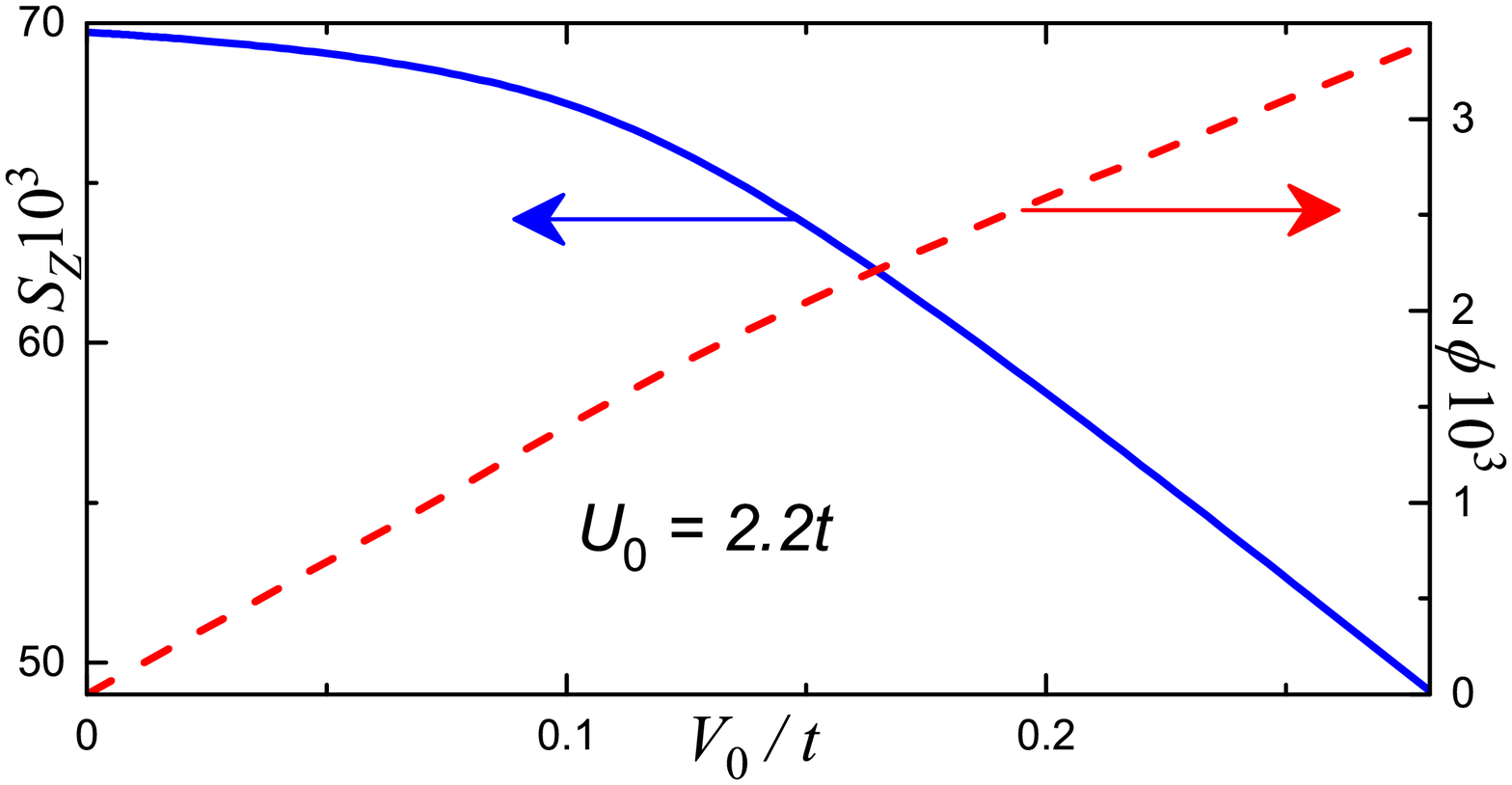}\\
\includegraphics[width=0.95\columnwidth]{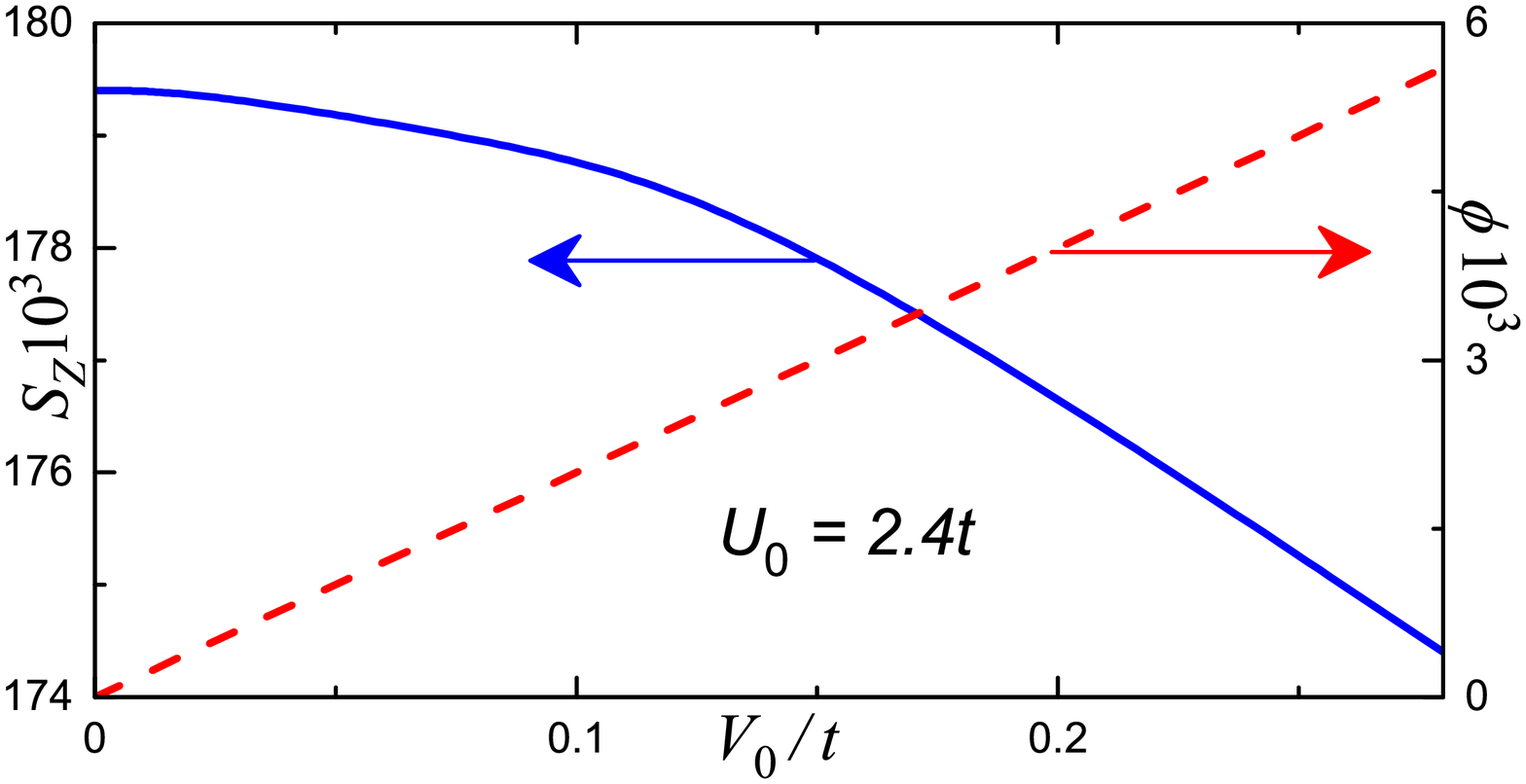}
\caption{(Color online) The dependence of the AFM magnetization $S_z$ and
of the exciton magnetization $\phi$ on the bias voltage $V_0$ for three
different values of the on-site interaction $U_0$. The blue continuous curves are
the AFM magnetization $S_z$, while the red dashed curves are for the
exciton magnetization $\phi$. For all panels we use the value
$U_{12}=U_0/4$.}\label{magnetizations}
\end{figure}

Let us analyze now the dependencies of the AFM and exciton order
parameters on the applied voltage. For the typical values of the
system parameters $U_0\simeq2.2t$, $U_{12}/U_0\simeq1/4$, and
$V_0/t_0\simeq1$, the values of the order parameters are
$\Delta_{\textrm{AFM}}\simeq0.17$\,eV and
$\Delta_{\textrm{exc}}\simeq8$\,meV. We can rewrite the expressions for the
order parameters in terms of magnetizations
\begin{eqnarray}
S_z&=&\frac{2\Delta_{\textrm{AFM}}}{U_0}=\langle n_{n1{\cal A}\uparrow} \rangle-\langle n_{n1{\cal A}\downarrow} \rangle,\\
\phi&=&\frac{2\Delta_{\textrm{exc}}}{U_{12}}=\langle d^{\dagger}_{n1{\cal A}\uparrow}d^{\phantom{\dagger}}_{n2{\cal A}\uparrow}\rangle-\langle d^{\dagger}_{n1{\cal A}\downarrow}d^{\phantom{\dagger}}_{n2{\cal A}\downarrow}\rangle.\nonumber
\end{eqnarray}
In these equations, the AFM magnetization $S_z$ is equal to the magnetization
per site of the sublattice
${\cal A}$
in layer 1. For the G-type AFM order, the magnetizations of electrons
located at neighboring sites have opposite signs.

The exciton magnetization $\phi$ can be viewed as the spin located on
the link connecting the sites $A1$ and $A2$ (for definitions of $A1$ and $A2$, see Fig. 1). The spin on the link connecting carbon atoms $B1$ and $B2$ has oppposite sign. The dependence of $S_z$
on the applied bias voltage calculated for three different values of the
on-site interaction constant $U_0$
is shown in Fig.~\ref{magnetizations}
by the solid lines. This magnetization is suppressed by the bias voltage. The
suppression is stronger for smaller $U_0$. When $U_0>2.4t$, the
magnetization $S_z$ only slightly depends on $V_0$. The exciton
magnetization $\phi$ is shown in Fig.~\ref{magnetizations} by dashed
lines. It increases almost linearly with $V_0$.
Nevertheless, $\phi$ is much smaller than $S_z$, even for relatively large
$V_0$.

\subsection{Exciton order parameter}

In the limit of small interactions $U_{12},U_0\ll t$, the second equation in~\eqref{ExcitonGap} simplifies and reduces to
\begin{eqnarray}
\Delta_{\textrm{exc}}=\Delta_{\textrm{AFM}}\frac{V}{2t_0}\frac{U_{12}}{U_0}.
\end{eqnarray}
In this limit the value of the exciton order parameter depends linearly on the inter-plane repulsion $U_{12}$.
References~\onlinecite{Joglekar,Min,Perali,Rossi,Sokolik,Efetov}
considered a system with two graphene layers separated by an insulating layer.
The dielectric barrier between the layers completely suppresses the interlayer
tunneling and destroys the AFM order. In this case, the value of the exciton
order parameter depends exponentially on the Coulomb interaction between
the layers. Under such conditions, according to Ref.~\onlinecite{Efetov}, the exciton gap becomes exponentially small around 1~mK.
In this case, a small amount of disorder makes exciton condensation
impossible~\cite{Das_Sarma}. In our
case, the exciton order parameter depends almost linearly on $U_{12}$.
Thus, the exciton order parameter can exist in our system even if the
inter-plane interaction is rather small.

Can we detect this order parameter? In principle, the exciton condensation can be observed experimentally by measuring the Coulomb
drag~\cite{How_to,Das_sarma_drag,Coulomb_drag}.
The experimental observation of Coulomb drag in bilayer graphene systems
with a dielectric barrier between the layers has been
reported~\cite{Gorbachev_drag}.
The execution and interpretation of similar experiment on bilayer graphene without the insulating layer might be a much more complicated issue.

All the above results were obtained at zero temperature. The detailed study of the temperature
dependence of the AFM order parameter at zero bias voltage was performed in
Ref.~\onlinecite{Last_paper}. Since the graphene bilayer is a two
dimensional system, it does not have a distinct magnetic phase transition. However,
we can define a crossover temperature $T^{*}$ between the short-range
antiferromagnetic and paramagnetic states. The calculations done in
Ref.~\onlinecite{Last_paper} show that
$T^{*}\approx0.5\Delta_{\text{AFM}}$. For realistic values of the applied
voltage the exciton order parameter is much smaller than the AFM order
parameter. Consequently,
$\Delta_{\text{exc}}\ll T^{*}$.
However the exciton order parameter is tied with the AFM order parameter,
and we expect that they both have the same crossover temperature about,
$T^{*}$
at
$V_0=0$.
Since the AFM order parameter can be high enough, the exciton order
parameter can survive at relatively high temperatures.

\section{Conclusions}
%%%%%%%%%%%%%%%%%%%%%%%%%%%%%%%%%%%%%%%%%%%%%%%%%%
In this paper we have studied theoretically the electronic properties of biased AA stacked bilayer graphene. The model Hamiltonian was analyzed
in the mean-field approximation. At zero bias, the ground state of the
system is antiferromagntic. We found that the applied transverse voltage
stabilizes the exciton order parameter coexisting with the AFM order. This
new order parameter couples the electrons and holes in different graphene
layers. The AFM phase with the coexisting exciton order parameter is the
most stable phase if the bias voltage is non-zero. The electronic gap is partially
suppressed by the bias voltage leading to a tunable metal-insulator
transition. The value of the exciton order parameter
can be about several tens of meV. Despite this small value, the exciton
order parameter can survive at relatively high temperatures due to its
coexistence with the AFM phase.

\section*{Acknowledgments}

This work was supported in part by RFBR (Grants
Nos.~14-02-00276, 14-02-00058, 12-02-00339), the RIKEN iTHES Project,
MURI Center for Dynamic Magneto-Optics, and a Grant-in-Aid for Scientific research (S).

\appendix

\section{Analytic solution}

Here we derive the analytical formulas for the AFM and exciton order
parameters.
We introduce dimensionless quantities $\delta_{\textrm{AFM}}=\Delta_{\textrm{AFM}}/t$ and $\delta_{\textrm{exc}}=\Delta_{\textrm{exc}}/t$.
It is convenient to rewrite the exciton order parameter in the following form
\begin{eqnarray}\label{qwe}
\delta_{\textrm{exc}}=\delta_{\textrm{AFM}}\frac{V}{2t_0}b\,,
\end{eqnarray}
where $b$ is the new variable.
Using this substitution we can rewrite Eqs.~\eqref{First_equation} and~\eqref{Second_equation} in the form
\begin{eqnarray}\label{EqDelta1}
\frac{4t}{U_0}&=&\!\!\!\int\limits_0^3\!\!d\zeta\,\rho_0(\zeta)\!\!
\left[\frac 1{\sqrt{\delta_{AFM}^2+(\zeta-\zeta_0)^2}}+\frac 1{\zeta+\zeta_0}\right]-\\
&&\!\!\!\!\!\!\!\!\!\!\int\limits_0^3\!\!d\zeta\,\rho_0(\zeta)\frac{V^2(1-b)}{4t^2\zeta_0\zeta}\!\!
\left[\frac 1{\sqrt{\delta_{AFM}^2+(\zeta-\zeta_0)^2}}-\frac 1{\zeta+\zeta_0}\right],\nonumber
\end{eqnarray}
\begin{eqnarray}\label{EqDelta2}
\frac{4t}{U_{12}}\!\!&=&\!\!\!\int\limits_0^3\!\!d\zeta\,\rho_0(\zeta)\!\!
\left[	\frac 1{\sqrt{\delta_{AFM}^2+(\zeta-\zeta_0)^2}}+\frac 1{\zeta+\zeta_0}\right]+\\
&&\!\!\!\!\!\!\!\!\!\!\int\limits_0^3\!\!d\zeta\,\rho_0(\zeta)\frac{t_0^2(1-b)}{t^2\zeta_0\zeta b}\!\!\left[\frac 1{\sqrt{\delta_{AFM}^2+(\zeta-\zeta_0)^2}}-\frac 1{\zeta+\zeta_0}\right],\nonumber
\end{eqnarray}
Taking the integration in Eq.~\eqref{EqDelta1} we obtain in the limit $\delta_{\textrm{AFM}}\ll1$
\begin{eqnarray}\label{Log}
\frac{4t}{U_0}&=&
\rho_0(\zeta_0)\left(1-\frac{V^2(1-b)}{4t^2\zeta_0^2}\right)\ln\left(\frac{4\zeta_0(3-\zeta_0)}{\delta_{\textrm{AFM}}^2}\right)+\nonumber\\ &&\eta_1(\zeta_0)+\frac{V^2(1-b)}{4t^2\zeta_0}\eta_2(\zeta_0)+O(\delta_{\textrm{AFM}}^2)\,,
\end{eqnarray}
where
\begin{eqnarray}\label{integrals}
\eta_1(\zeta_0)\!&=&\!\int\limits_0^3\!\!d
\zeta\left[		\frac{\rho_0(\zeta)}{\zeta+\zeta_0}
		+		\frac{\rho_0(\zeta)-\rho_0(\zeta_0)}
		     {\left|\zeta-\zeta_0\right|}\right],\;\;\
\nonumber \\
\eta_2(\zeta_0)\!&=&\!\int\limits_0^3\!\!\frac{d\zeta}{\zeta}\left[		 \frac{\rho_0(\zeta)}{\zeta+\zeta_0}
		-		\frac{\rho_0(\zeta)\zeta_0-\rho_0(\zeta_0)\zeta}
		     {\zeta_0\left|\zeta-\zeta_0\right|}\right].\;\;\
\end{eqnarray}
Performing the similar integration in Eq.~\eqref{EqDelta2} and expressing the logarithmic term using Eq.~\eqref{Log} we obtain in the limit of small $b\ll 1$ the following equation for $b$
\begin{equation}\label{b}
b\!=\!\frac{\frac{4t}{U_0}-\eta_1(\zeta_0)-\eta_2(\zeta_0)\zeta_0}{\frac{4t}{U_{12}}\!-\!\frac{4t}{U_0}\frac{t^2\zeta_0^2}{t_0^2}\!+\!\frac{V^2(\eta_1(\zeta_0)\!+\!\eta_2(\zeta_0)\zeta_0)}{4t_0^2}}.
\end{equation}
For the range of parameters $U_0$ and $U_{12}$ under study, we have $b<0.05$, so the assumption $b\ll1$ is well satisfied. The expression for $\Delta_{\textrm{exc}}$ is written as follows
\begin{equation}
\Delta_{\textrm{exc}}\!=\!\Delta_{\textrm{AFM}}\!\frac{V}{2t_0}\frac{\frac{4t}{U_0}-\eta_1(\zeta_0)-\eta_2(\zeta_0)\zeta_0}{\frac{4t}{U_{12}}\!-\!\frac{4t}{U_0}\frac{t^2\zeta_0^2}{t_0^2}\!+\!\frac{V^2(\eta_1(\zeta_0)\!+\!\eta_2(\zeta_0)\zeta_0)}{4t_0^2}}\,.
\end{equation}
The antiferromagnetic gap is found from Eq.~\eqref{Log}, where we can neglect $b$ in the first and third terms. As a result, we obtain
\begin{equation}
\Delta_{\textrm{AFM}}\!=\!2t\!\sqrt{\!\zeta_0(\!3\!-\!\zeta_0\!)}\!\exp\!\left(\!\frac{\eta_1(\zeta_0)+\frac{\eta_2(\zeta_0)V^2}{4t^2\zeta_0^2}-\frac{4t}{U_0}}{2\rho_0(\zeta_0)\frac{t_0^2}{t^2\zeta_0^2}}\!\right)\!.
\end{equation}
\end{document}